\documentclass[aps,prb,onecolumn,preprint,superscriptaddress]{revtex4}
\usepackage{amssymb}

\usepackage{graphicx}

\begin{document}

\title{Valley Trion Dynamics in Monolayer MoSe$_{2}$}

\author{Feng Gao}
\affiliation{
     Department of Physics, Florida International University, Miami, Florida 33199
}

\author{Yongji Gong}
\affiliation{
     Department of Chemistry, Rice University, Houston, Texas 77005
}

\affiliation{
     Department of Materials Science and NanoEngineering, Rice University, Houston, Texas 77005
}

\author{Michael Titze}
\affiliation{
     Department of Physics, Florida International University, Miami, Florida 33199
}

\author{Raybel Almeida}
\affiliation{
     Department of Physics, Florida International University, Miami, Florida 33199
}

\author{Pulickel M. Ajayan}
\affiliation{
     Department of Chemistry, Rice University, Houston, Texas 77005
}

\affiliation{
     Department of Materials Science and NanoEngineering, Rice University, Houston, Texas 77005
}

\author{Hebin Li}
\affiliation{
     Department of Physics, Florida International University, Miami, Florida 33199
}


\date{\today}

\begin{abstract}
Charged excitons called trions play an important role in the fundamental valley dynamics in the newly emerging 2D semiconductor materials. We used ultrafast pump-probe spectroscopy to study the valley trion dynamics in a MoSe$_2$ monolayer grown by using chemical vapor deposition. The dynamics display an ultrafast trion formation followed by a non-exponential decay. The measurements at different pump fluences show that the trion decay dynamics become slower as the excitation density increases. The observed trion dynamics and the associated density dependence are a result of the trapping by two defect states as being the dominating decay mechanism. The simulation based on a set of rate equations reproduces the experimental data for different pump fluences. Our results reveal the important trion dynamics and identify the trapping by defect states as the primary trion decay mechanism in monolayer MoSe$_2$ under the excitation densities used in our experiment.
\end{abstract}


\maketitle
\section{Introduction}

Layered transition metal dichalcogenides (TMD), $MX_2$ ($M=Mo, W; X=S, Se, Te$), represent a new class of atomically thin two-dimensional (2D) materials inspired by the discovery of graphene \cite{Novoselov2005}. In contrast to graphene, monolayer TMD has a direct band gap \cite{Jin2013,Zhang2014} in the visible region and displays strong photoluminescence (PL) at the $K$ and $-K$ points in the Brillouin zone \cite{Mak2010,Splendiani2010}. Moreover, the inversion symmetry breaking and strong spin-orbit coupling in monolayer TMD lead to contrasting circular dichroism in the $\pm K$ valleys \cite{Xiao2012}. Consequently, the interband transitions at the two valleys can be selectively excited by an optical field with proper helicity. This ability to selectively address different valleys enables optical generation, control and detection of valley polarizations \cite{Cao2012,Mak2012,Zeng2012,Sallen2012} and valley coherence\cite{Jones2013}. These unique properties and recent advances in sample fabrication \cite{Wang2014b} make layered TMD promising materials for novel applications in optoelectronics\cite{Wang2012} and the field of valleytronics\cite{Xu2014, Shkolnikov2002, Bishop2007, Yao2008, Xiao2007, Rycerz2007, Gunawan2006}.

The optical excitation of semiconductors creates excitons which are electron-hole pairs bound through Coulomb interactions. In the presence of excess charges, charged excitons called trions \cite{Lampert1958,Kheng1993} can be formed through the coalescence of an exciton and a free charge or directly from an unbound electron-hole plasma \cite{Portella-Oberli2009}. In monolayer TMD, strong Coulomb interactions lead to exceptionally high binding energies for excitons \cite{Thilagam1997} and trions \cite{Jones2013, Mak2013, Ross2013}, allowing them to exist even at room temperature. Like excitons, trions also have significant influences on the optical and electronic properties of layered TMD. For example, trions can modify the overall PL spectrum \cite{Ross2013} in monolayer MoSe$_2$ and reduce the conductivity \cite{Lui2014} in monolayer MoS$_2$. The interplay between exciton and trion is critical. The trion state has a lower energy and provides a relaxation channel for excitons. Trions can be excited by an optical phonon into an excitonic state to realize a upconversion process \cite{Jones2015} in monolayer WSe$_2$. Coherent exciton-trion coupling has also been observed \cite{Singh2014} in monolayer MoSe$_2$. Moreover, trions play an important role in applications such as quantum information processing \cite{Xu2008}. Therefore, a thorough understanding of the fundamental carrier dynamics in monolayer TMD has to include both exciton and trion dynamics.

The experimental investigation of the carrier dynamics upon an optical excitation has been the focus of recent studies based on techniques such as time-resolved PL \cite{Korn2011, Lagarde2014a}, ultrafast pump-probe spectroscopy\cite{Shi2013a, Wang2013b, Cui2014, Kumar2014, Mai2014, PhysRevB.90.041414, Mannebach2014, Sun2014, Wang2015, PhysRevLett.115.257403, PhysRevB.93.041401}, optical Kerr spectroscopy\cite{Yang2015,Yang2015a}, coherent 2D spectroscopy\cite{Moody2015} and hole-burning spectroscopy\cite{Schaibley2015}. These studies have revealed valuable structure and dynamic information in layered TMD. However, the focus has been mainly on the exciton dynamics while the trion dynamics remain largely unexplored. Recent studies of the trion formation in monolayer MoSe$_2$ by pump-probe spectroscopy \cite{PhysRevB.93.041401} and the trion emission in monolayer WSe$_2$ by time-resolved PL \cite{Wang2014a} have brought the attention to the valley trion dynamics in layered TMD.

Here we report a study of the valley trion dynamics in a chemical-vapor-deposition (CVD) grown MoSe$_2$ monolayer by using ultrafast pump-probe spectroscopy. We find that the trion population decays non-exponentially after an ultrafast ($<$ 500 fs) trion formation from photoexcited free carriers. The trion decay can be fit with a bi-exponential decay function. The measurements at different pump fluences show a surprising density dependence of the trion decay: the dynamics become slower as the excitation density increases. We present a theoretical model based on a set of rate equations that reproduces the experimental data quantitatively for all pump fluences. The model shows that the primary mechanism responsible for the observed dynamics and density dependence is the fast and slow trapping by two defect states. The slower dynamics at higher densities is due to the limited density of the defects.

\begin{figure}[htbp]
 \centering
  \includegraphics[width=\columnwidth]{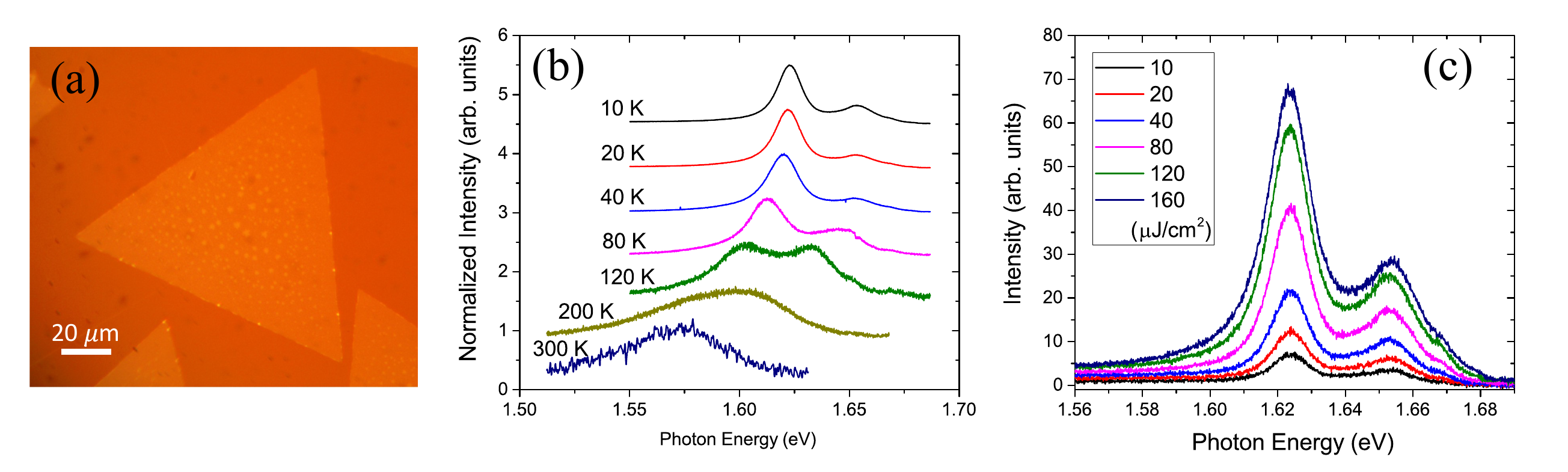}
\caption{(a) The optical image of a CVD MoSe$_2$ monolayer. (b) Normalized PL spectra at various temperatures from 10 to 300 K. The spectra are shifted vertically for clarity. (c) PL spectra at 10K with various pump fluences from 10 to 160 $\mu$J/cm$^2$.}
\label{figure1}
\end{figure}

\section{Results and discussion}

The CVD-grown MoSe$_2$ monolayers \cite{Wang2014} on a glass substrate were studied. The optical image of a typical monolayer MoSe$_2$ flake is shown in Fig. \ref{figure1}(a). The flakes are equilateral triangles with sides ranging from tens of $\mu$m to more than 100 $\mu$m. The sample was placed in a microscopy liquid helium cryostat so that the sample temperature can be varied from 10 K to the room temperature. We first performed PL measurements at different temperatures. Femtosecond pulses with the central wavelength at 720 nm were used as the pump for both the PL and pump-probe experiments. The PL spectra at various temperatures from 10 K to 300 K are shown in Figure \ref{figure1}(b). As the temperature decreases, the PL spectrum changes from a broad single peak to two separate peaks with an energy difference of $\sim$30 meV. At low temperatures, the peak at the higher energy (1653 meV at 10K) is identified as the $A$ exciton resonance while the one at the lower energy (1623 meV at 10K) as the trion resonance. This assignment is consistent with previously published studies \cite{Ross2013, PhysRevB.93.041401, Singh2014} in the PL spectra, the resonance energies and the trion binding energy. The PL was also measured under different pump fluences to check how the excitation density affects the exciton and trion resonances. Figure \ref{figure1}(c) shows the PL spectra at 10 K under various pump fluences. The spectra are fit to a Lorentzian double peak profile to extract peak positions, linewidths, and peak areas (see Supporting Information for details). As the pump fluence changes from 10 to 160 $\mu$J/cm$^2$, the exciton and trion resonance energies have a small variation of 1.5 and 0.4 meV, respectively. The linewidth increases from 15.6 to 18.1 meV for exciton and from 13.1 to 16.65 meV for trion. While both exciton and trion intensities increase linearly with the pump fluence, the ratio of exciton to trion peak area decreases from 0.52 to 0.38. This ratio can be used to determine the distribution of the initial carrier density between the exciton and trion states.

\begin{figure*}[bt]
 \centering
  \includegraphics[width=\textwidth]{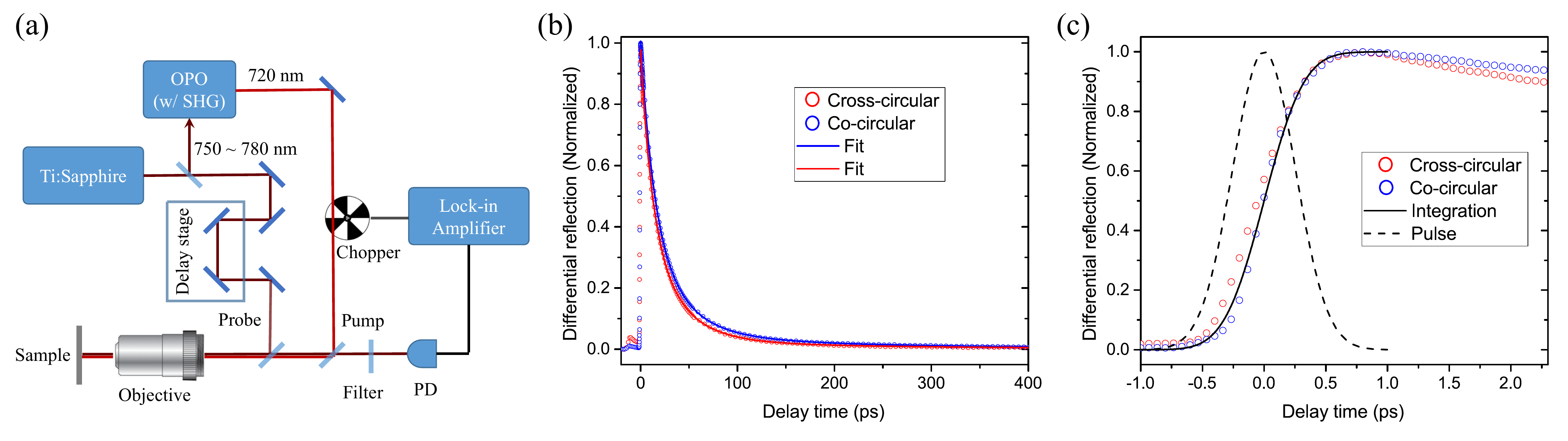}
\caption{(a) The schematic of the pump-probe experiment. (b) Normalized pump-probe spectra with cross-circular (red circles) and co-circular (blue circles) polarizations at 10 K with a pump fluence of 80 $\mu$J/cm$^{2}$. The solid lines are fit to a bi-exponential decay function. (c) A zoom-in showing the rising slope. The dash line is a Gaussian pulse with a duration (FWHM) of 500 fs. The solid line is the normalized integration of the Gaussian pulse.}
\label{figure2}
\end{figure*}

The ultrafast pump-probe experiment was performed with the setup shown in Figure \ref{figure2}(a). The pump pulse was tuned to 1722 meV to excite free carriers slightly above the $A$ exciton energy but below the $B$ exciton energy. The probe pulse was tuned to probe at the peak of the trion resonance at 1623 meV. The bandwidth of the probe pulse ($\sim$9 meV) is much larger than the variation in the trion resonance energy. The pump and probe beams are circularly polarized to selectively excite and probe the transition at one of the $\pm K$ valleys depending on the helicity. Representative pump-probe spectra are shown in Figure \ref{figure2}(b) for a sample temperature of 10 K and a pump fluence of 80 $\mu$J/cm$^2$. The blue (red) circles are the data obtained with the co-circularly (cross-circularly) polarized pump and probe beams, that is, the pump and probe polarizations are $\sigma^+\sigma^+$ ($\sigma^+\sigma^-$). The pump-probe spectra feature a fast rising followed by a decay of the signal up to hundreds of ps.

To investigate the rising signal, Figure \ref{figure2}(c) shows a segment of several ps around zero time delay. In addition to the pump-probe spectra, a 500-fs Gaussian pulse is plotted as a dash line. The normalized time integral of the Gaussian pulse, shown as a solid line, matches well with the rising slope of the pump-probe spectra. Therefore, the rising time of the signal is limited only by the time resolution of the pump pulse. For the co-circular configuration, this indicates that the trion formation time is within 500 fs. This time is consistent with the trion formation time and its dependence on the excitation energy reported by Singh and coauthors \cite{PhysRevB.93.041401}. The variation in the trion formation time can be caused by the differences in doping density, excitation energy and excitation power.
The spectrum with the cross-circular configuration shows a nearly identical rising signal, suggesting a fast intervalley relaxation time. This seems contradictory to the observations of a persistent polarization in the PL of excitons \cite{Mak2012,Zeng2012,Sallen2012,Lagarde2014a} and trions \cite{Wang2014a}. However, the reported circular polarization degree is considerably less than 100\% in most experiments. The imperfect but persistent polarization can be explained with an ultrafast initial polarization decay due to fast intervalley relaxation, in which some carriers lose the initial polarization quickly while others maintain for a longer time. The partial loss of polarization is also evident by our observation that the amplitude of the pump-probe signal, prior to the normalization, with the cross-circular polarizations is about 30\% of the amplitude with the co-circular polarizations. The reduced trion intensity for the cross-circular polarizations has also been observed in the PL\cite{Wang2014a} spectra. The time-resolved PL measurements \cite{Lagarde2014a, Wang2014a} suggested an intervalley relaxation time shorter than the time resolution of 4 ps in their experiments. Our data show that the intervalley relaxation time is shorter than 500 fs and only a part of carrier population is involved in this ultrafast initial intervalley relaxation.

It has been shown \cite{Portella-Oberli2009} that the trion formation has two possible channels. In the presence of excess free charges, an exciton can capture an extra charge to form a trion. This is the primary process for trion formation at low carrier densities. At a sufficiently high density, trions can be formed directly from unbound carriers through a three-particle formation process. In our experiment, the trion formation is as fast as the exciton formation within the time resolution of our measurement. We suspect that the trion formation at the short time scale is mainly the three-particle formation from free carriers. However, our results cannot rule out the possibility of an ultrafast ($<$ 500 fs) exciton-to-trion formation process. As we will show in the simulation, a slow trion formation process through the coalescence of excitons and excess charges may exist and lead to a better agreement between the simulated and experimental data at long delay times.

\begin{figure*}[htbp]
 \centering
  \includegraphics[width=\textwidth]{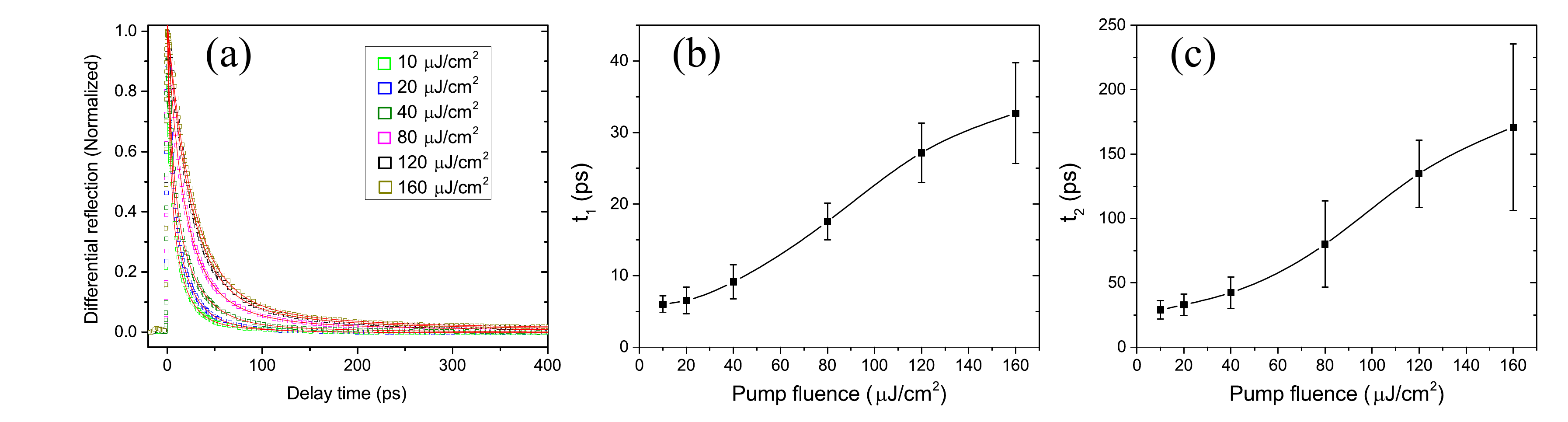}
\caption{(a) Normalized pump-probe spectra obtained at 10 K with the pump fluences ranging from 10 to 160 $\mu$J/cm$^2$. The solid lines are fits to the bi-exponential decay function. (b) The extracted values of the time constant $t_1$ at different pump fluences. (c) The extracted values of the time constat $t_2$ at different pump fluences.}
\label{figure3}
\end{figure*}

After the rise, the pump-probe signal displays a non-exponential decay which includes a fast decay at the time scale within tens of ps and a slower decay at hundreds of ps. Similar behaviors in the decay signal have been observed for both excitons \cite{Cui2014, Korn2011, Kumar2014, Lagarde2014a, Wang2015, PhysRevB.90.041414, Shi2013a, PhysRevLett.115.257403} and trions \cite{Wang2014a} in various TMD monolayers. A bi-exponential decay function provides a good fit to the data in some cases\cite{Cui2014,Korn2011,Kumar2014,Lagarde2014a,Wang2015} while other experiments\cite{PhysRevB.90.041414,Shi2013a,PhysRevLett.115.257403} suggest that a tri-exponential decay function is required. In our experiments, a bi-exponential decay function is sufficient to fit the decay signal. To analyze the decay dynamics, we use the bi-exponential decay function, $y=A_0(e^{-t/t_1}+A_{21}e^{-t/t_2})$, to fit the data. The solid lines in Fig. \ref{figure2}(b) are fits with the fitting parameters $t_1=18.4\pm0.2$ ps, $t_2=80\pm2$ ps, and $A_{21}=0.19\pm0.01$ for the co-circular polarizations, and $t_1=16.8\pm0.3$ ps, $t_2=66\pm2$ ps, and $A_{21}=0.20\pm0.02$ for the cross-circular polarizations. The fit to the bi-exponential decay function allows a global characterization of the decay time constants for all data, however it does not necessarily suggest that there are only two relaxation processes or attribute the time constants to a particular process.

A number of exciton decay mechanisms have been suggested as possible underlying processes in previously published studies\cite{Korn2011, Lagarde2014a, Shi2013a, Wang2013b, Cui2014, Kumar2014, Mai2014, PhysRevB.90.041414, Mannebach2014, Sun2014, Wang2015, PhysRevLett.115.257403, PhysRevB.93.041401}. For instance, some experiments\cite{Kumar2014,Sun2014} show that the fast decay dynamics ($< 50$ ps) are dependent on the exciton density and exciton-exciton annihilation is the dominating decay channel, while other studies\cite{Shi2013a,Wang2015,Wang2013b} show no significant variation in the decay dynamics with the excitation density and that trapping by surface defect states is the responsible process. Other mechanisms such as carrier-phonon scattering, inter/intra-valley scattering, biexciton formation, trion formation, electron-hole recombination and exciton Auger scattering have also been considered \cite{Moody2016}. Some of the exciton decay mechanisms can also be applicable to the trion dynamics. For instance, trapping by shallow and deep defect states can lead to the bi-exponential decay observed in the trion dynamics. The studies on exciton dynamics \cite{Korn2011, Lagarde2014a, Shi2013a, Wang2013b, Cui2014, Kumar2014, Mai2014, PhysRevB.90.041414, Mannebach2014, Sun2014, Wang2015, PhysRevLett.115.257403, PhysRevB.93.041401} have shown considerable discrepancy in the interpretation of the exciton dynamics and the specific time scales due to the differences in samples and measurement conditions such as the excitation energy, excitation density, temperature and polarizations. Particularly, the dependence on the excitation density can provide crucial evidence in determining the decay mechanism.

To further understand the trion relaxation dynamics in our sample, pump-probe spectra were obtained with the pump fluence varying from 10 to 160 $\mu$J/cm$^2$ to investigate the excitation density dependence. The injected carrier density is estimated to be $1.8\times 10^{12}$ cm$^{-2}$ for a pump fluence of 10 $\mu$J/cm$^2$ by using an absorption measurement and assuming that one electron is excited into the conduction band for each absorbed photon. The normalized pump-probe spectra at 10 K with various pump fluences are shown in Figure \ref{figure3}(a). The squares with different colors are experimental data and solid lines are fits to the bi-exponential decay function. The decay dynamics show a strong dependence on the pump fluence, thus the initial excitation density. The extracted time constants $t_1$ and $t_2$ are plotted in Figure \ref{figure3}(b) and (c), respectively, as a function of the pump fluence. The error bars are estimated from the fitting uncertainties and multiple measurements. Both time constants increase with the pump fluence, indicating that the trion decay is slower at higher excitation densities. This dependence is surprising since the dynamics is usually expected to be faster due to stronger many-body interactions as the density increases.


For exciton dynamics, the excitation-density dependence can be a result of different processes such as biexciton formation \cite{You2015, Plechinger2015}, exciton-exciton annihilation \cite{Kumar2014,Sun2014} and defect-assisted Auger scattering \cite{Wang2015,Wang2015a}. Similar processes may contribute to the trion dynamics. We rule out the biexciton formation since there is no evidence of biexciton in monolayer MoSe$_2$ despite the observation of biexcitons in WSe$_2$ \cite{You2015} and WS$_2$ \cite{Plechinger2015}. In a possible trion-trion annihilation process, the trion density should decay quadratically according to the rate equation $dN/dt=-kN^2$ with $N$ being the trion density and $k$ the annihilation rate. The decay governed by this equation should become faster as the initial density increases, opposite to the excitation-density dependence of the dynamics in our experiment. Therefore, we also rule out the trion-trion annihilation process.

\begin{figure}[htbp]
 \centering
  \includegraphics[width=\columnwidth]{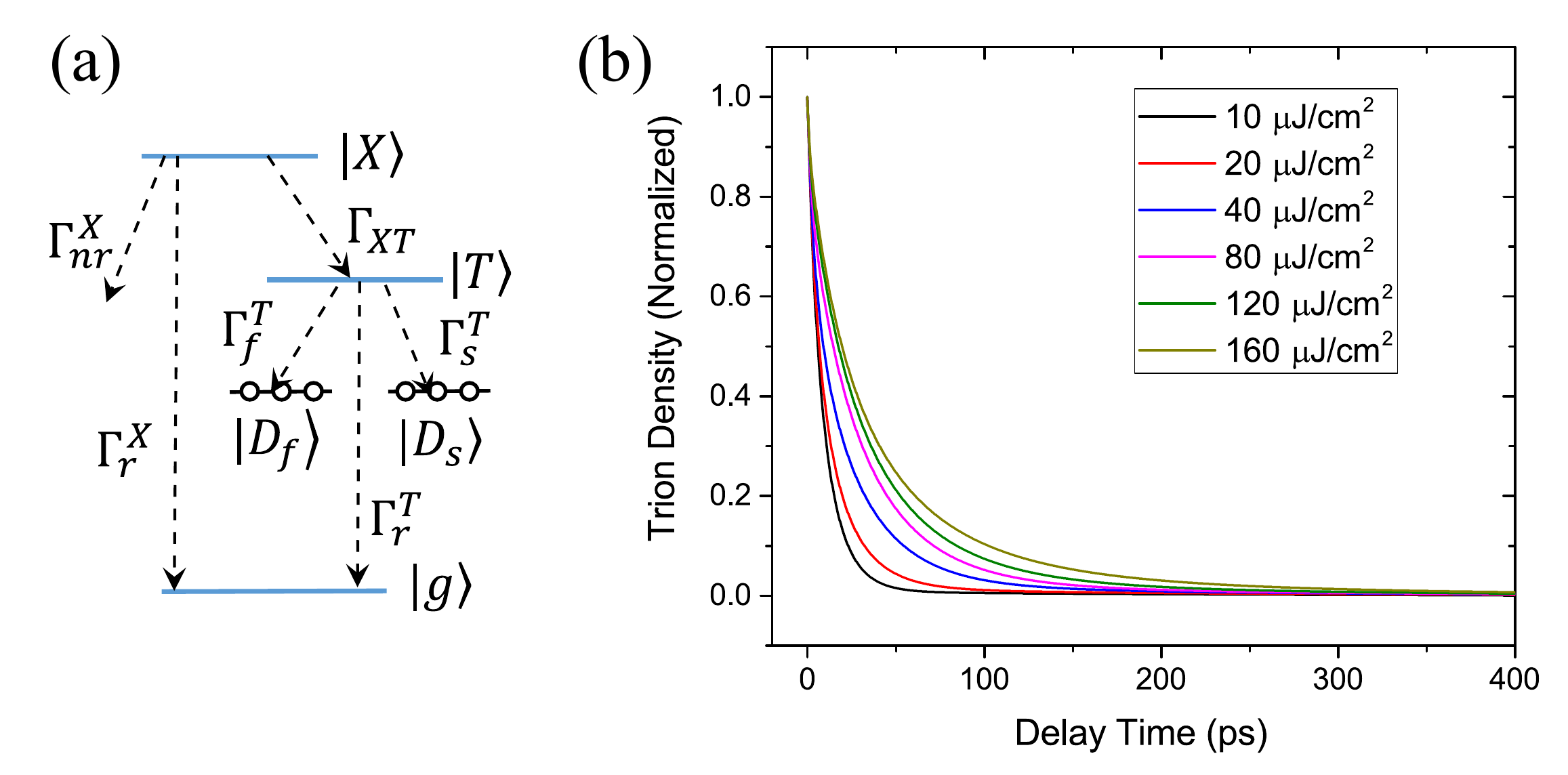}
\caption{(a) The energy-level diagram showing the exciton and trion decay channels. (b) The simulated trion decay dynamics for different pump fluences.}
\label{figure4}
\end{figure}

The bi-exponential decay and the excitation-density dependence of the trion dynamics can be explained by the trapping of trions in the defect states through Auger scattering. Similar to the trapping of carriers and excitons, a trion can be captured into a defect state through either a phonon-assisted process or an Auger process. With phonon-assisted processes, the trapping rates depend strongly on the temperature. Our measurements at different temperatures (see Supporting Information) do not show such a temperature dependence, suggesting that phonon-assisted processes are negligible in our case. In an Auger scattering process, an electron (hole) scatters off a hole (electron) and is captured into a defect state while the hole (electron) is scattered to a higher energy state to conserve energy. The corresponding exciton or trion is trapped in the defect state. The defect trapping has been experimentally studied in MoS$_2$ monolayers \cite{Wang2015} and nanoclusters \cite{Doolen1998}. Two defect states have been observed to provide both fast and slow traps, which can explain the bi-exponential decay. In MoS$_2$ monolayers, the decay dynamics were found \cite{Wang2015} to be independent of the excitation density within a range of the pump fluence from 1 to 32 $\mu$J/cm$^2$. However, the density dependence of the dynamics in our experiment can be explained by considering the limited density of the available defect states in the sample. As the exciton/trion density approaches or becomes higher than the defect density, the defect states are filled up and the defect trapping process slows down. If this is the dominating decay channel, the trion relaxation becomes slower as the initial density increases.

To test this hypothesis, we model the trion relaxation dynamics with rate equations based on the energy level structure shown in Figure \ref{figure4}(a). The exciton, trion and ground states are considered in the excitonic picture. The excitons decay through both radiative and non-radiative channels at the rates $\Gamma_r^X$ and $\Gamma_{nr}^X$, respectively. The excitons can also capture an extra charge to form trions at the rate $\Gamma_{XT}$. The trions decay through the radiative recombination at the rate $\Gamma_r^T$ and the defect trapping. We assume that there are two defect states providing fast trapping at the rate $\Gamma^T_f$ and slow trapping at the rate $\Gamma^T_s$. The dynamics can be described by a set of rate equations

\begin{eqnarray}
\dot{N}_{X1}&=&-\Gamma_{nr}^X N_{X1}; \\
\dot{N}_{X2}&=&-\Gamma_r^X N_{X2}-\Gamma_{XT} (1-\frac{N_T}{D_T}) N_{X2}; \\
\dot{N}_T&=& -\Gamma_r^T N_T - \Gamma_f^T (1-\frac{N_f}{D_f}) N_T -  \Gamma_s^T (1-\frac{N_s}{D_s}) N_T + \Gamma_{XT} N_{X2};\\
\dot{N}_f &=& \Gamma_f^T(1-\frac{N_f}{D_f})N_T; \\
\dot{N}_s &=& \Gamma_s^T(1-\frac{N_s}{D_s})N_T;
\end{eqnarray}
where $N_{X1}$ ($N_{X2}$) is the exciton density undergoing the fast (slow) decay, $N_T$ is the trion density, $N_f$ ($N_s$) is the occupied density in the fast (slow) state, $D_f$ ($D_s$) is the defect density for the fast (slow) trapping, and $D_T$ is the density of the available trion states. The photoexcited carriers form excitons and trions in a time shorter than the pulse duration. The initial exciton and trion densities are estimated from the injected carrier density and the ratio of exciton to trion density obtained from the PL spectra with the corresponding pump fluence (see Supporting Information). The excitons have a fast decay channel due to non-radiative processes and a slow radiative decay channel. The excitons also form trions at a time scale of hundreds of ps. The trion-to-exciton upconversion \cite{Jones2015} has a low efficiency and is negligible in our experiment. We assume that $50\%\sim90\%$ of the initial exciton population decays through the fast channel and the remaining through the slow channel. The simulation shows that the choice of this ratio does not qualitatively affect the trion dynamics. The rate equations are solved numerically for the trion density as a function of time. The parameters are adjusted by a nonlinear fitting routine to fit all six decay curves in Figure \ref{figure3}(a) simultaneously. The simulated trion decay dynamics under different pump fluences are presented in Figure \ref{figure4}(b). The corresponding fitting parameters are shown in Table \ref{tbl:table1}.

The simulated result is in a good agreement with the experimental data. The simulation reproduces the non-exponential decay and the decay dynamics become slower as the pump fluence increases. We find that the density dependence of the trion dynamics is mostly affected by the defect densities $D_f$ and $D_s$. The best fitting requires $D_f$ ($D_s$) to be close to the lowest (highest) injected carrier density. The obtained defect densities are comparable with the theoretical and experimental values \cite{Komsa2012a,Zande2013,Liu2013,Zhou2013} of the point defects in 2D TMD. Although the trapping rates, $\Gamma^T_f$ and $\Gamma^T_s$, stay the same for all pump fluences, the actual decay times are longer at higher densities since the defect states are filled up and the effective trapping rates, $\Gamma^T_{f/s}(1-N_{f/s}/D_{f/s})$, decrease with the occupied density. The fast and slow trapping by the defect states explains the non-exponential decay and the density dependence in the experimental results, but slightly overestimates the decay rate at long delay times ($>$ 100 ps) for high pump fluences ($\geq$ 80 $\mu$J/cm$^2$). Adding a slow process of trion formation from excitons at the rate $\Gamma_{XT}$ provides a better fit to the decay dynamics at long delay times. However, this process is not essential to reproducing the density dependence of the trion dynamics. Therefore, the trion decay dynamics is dominated by the fast and slow trapping by the defect states within the range of the pump fluences used in our experiment.

\begin{table}
  \caption{The values of the parameters used in the simulation to fit the experimental data.}
  \label{tbl:table1}
\begin{tabular}{|l|r|}

  \hline
  Parameter & Value \\
  \hline
  $1/\Gamma^T_f$ & 10.3 ps \\
  \hline
  $1/\Gamma^T_s$ & 28.2 ps \\
  \hline
  $1/\Gamma_{XT}$ & 349 ps \\
  \hline
  $1/\Gamma^T_r$ & 500 ps \\
  \hline
  $1/\Gamma^X_r$ & 400 ps \\
  \hline
  $D_f$ & $1.60\times 10^{12}$ cm$^{-2}$ \\
  \hline
  $D_s$ & $3.06\times 10^{13}$ cm$^{-2}$ \\
  \hline
  $D_X$ & $5\times 10^{13}$ cm$^{-2}$ \\
  \hline 
\end{tabular}
\end{table}

\section{Conclusion}

We have studied the trion dynamics in monolayer MoSe$_2$ at a temperature of 10 K by using ultrafast pump-probe spectroscopy. The time scale of the trion formation from photoexcited free carriers is within 500 fs. The trion population decays non-exponentially and the decay signals can be fit with a bi-exponential decay function with a fast and slow time constant. Both time constants increase as the pump fluence increases, indicating slower trion decay dynamics at higher excitation densities. The non-exponential decay dynamics and their unconventional density dependence are attributed to the trapping by two defect states assisted by Auger scattering. The simulation based on the rate equations describing the trapping by defects and the exciton-to-trion formation process reproduces the measured trion decay dynamics for all pump fluences. The slower dynamics at higher densities are due to the limited defect densities that are comparable to the excitation densities. Our experiment and simulation suggest that the dominating mechanism for the trion decay dynamics is the trapping by two defect states at low temperatures within the range of the pump fluences used in our experiment.

\section{Experimental Methods}

Monolayer MoSe$_2$ was synthesized by using the chemical vapor deposition process as described in detail in a previous report \cite{Wang2014b}. Monolayer flakes were grown on a SiO$_2$/Si substrate and transferred to a glass substrate. The flakes are confirmed to be monolayers by Raman and PL spectroscopy. The sample was kept in a microscopy liquid helium cryostat (Cryo Industries of America RC102-CFM) for low temperature experiments.

Ultrafast pump-probe spectroscopy was performed with a Ti:sapphire femtosecond oscillator (Coherent Mira 900) and an optical parametric oscillator (OPO) (Coherent Mira OPO), as shown in Fig. \ref{figure2}(a). The femtosecond oscillator output is split to pump the OPO and work as the probe. The central wavelength of the femtosecond oscillator output is tuned to match the central wavelength of the trion PL peak. The OPO output works as the pump and the central wavelength is tuned to 720 nm (1.722 meV). The pulse duration is $\sim$500 fs ($\sim$200 fs) for the pump (probe) and the repetition rate is 76 MHz. The pump and probe beams are focused on the sample by a long working distance 50$\times$, 0.55 NA objective lens (Mitutoyo Plan APO 50$\times$). On the sample, the pump has a spot size of 7 $\mu$m in diameter and the probe spot is 5 $\mu$m in diameter. The probe fluence is 1 $\mu$J/cm$^{2}$ and the pump fluence is varied from 10 to 160 $\mu$J/cm$^{2}$. We tested the probe fluence of 2 $\mu$J/cm$^{2}$ and found no noticeable difference in the measured dynamics. The pump and probe polarizations can be controlled independently. The reflected probe beam is collected by the same objective lens and recorded with a photodetector and a lock-in amplifier while the pump beam is modulated by a chopper. The time delay of the probe pulse can be varied by a delay stage. A pump-probe spectrum is acquired by measuring the reflected probe as a function of the time delay between the pump and probe pulses.

The PL measurements were done in the same setup with the probe beam being blocked. The PL signal was collected by a half-meter spectrometer (Horiba iHR550) and recorded on a TEC cooled CCD camera (Horiba SYN-2048X512). The PL spectra were obtained with various pump fluences at different sample temperatures.

\end{document}